\documentclass[prl,twocolumn,showpacs,amsmath,floatfix,superscriptaddress,amssymb]{revtex4}
\usepackage[dvips]{graphicx}
\usepackage{dcolumn}
\usepackage{wrapfig}
\usepackage[small,compact]{titlesec}

\def\gapx{\ \lower 2pt \hbox{$\buildrel>\over{\scriptstyle{\sim}}$}}
\def\lapx{\ \lower 2pt \hbox{$\buildrel<\over{\scriptstyle{\sim}}$}}
\def\lap{\ \lower 2pt \hbox{$\buildrel<\over{\scriptstyle{\sim}}$}}
\def\no{$n_0$}
\def\nstark{$n^\star({\bf k})$}
\def\noft{$n_0(T)$}
\def\tl{$T_\lambda$}
\def\2d{$2\Delta$}
\def\3he{$^3$He}
\def\4he{$^4$He}

\def\a1{$a_{1}$}
\def\a2{$\overline\alpha_{2}$}
\def\alf{$\overline{\alpha}_4$}

\def\alt{$\overline{\alpha}_2$}
\def\als{$\overline{\alpha}_6$}

\def\Am3{\AA$^{-3}$}
\def\ax2{$a_{2}$}

\def\d2o{D$_2$O}

\def\h2o{H$_2$O}

\def\ke{$\langle{K}\rangle$}
\def\Ke3{$\langle{K_3}\rangle$}
\def\k2{$\langle{[k_{\alpha}]^2}\rangle$}

\begin{document}

\title{ Bose-Einstein Condensation in liquid $^4$He near the liquid-solid transition line}
\author{S.O. Diallo}
\affiliation{Spallation Neutron Source, Oak Ridge National Laboratory, Oak Ridge, TN 37831-6477, USA}
\author{R.T. Azuah}
\affiliation{NIST center for Neutron Research, Gaithersburg, MD 20742-2115, USA}
\affiliation{Department of Materials Science and Engineering, University of Maryland, College Park, MD  20742-2115, USA}
\author{D.L. Abernathy}
\affiliation{Spallation Neutron Source, Oak Ridge National Laboratory, Oak Ridge, TN 37831-6477, USA}
\author{R. Rota}
\affiliation{Departament de F\'isica i Enginyeria Nuclear, Universitat Polite\'ecnica de Catalunya, E-08034, Barcelona, Spain}
\author{J. Boronat}
\affiliation{Departament de F\'isica i Enginyeria Nuclear, Universitat Polite\'ecnica de Catalunya, E-08034, Barcelona, Spain}
\author{H.R. Glyde}
\affiliation{University of Delaware, Newark, DE, USA 19716-2570, USA}

\pacs{03.75.Kk, 78.70.Nx,67.80.bd}

\begin{abstract}
We present precision neutron scattering measurements of the Bose-Einstein condensate fraction, $n_0(T)$, and the atomic momentum distribution, $n^{\star}({\bf k})$, of liquid $^4$He at pressure $p=$24 bar.  Both the temperature dependence of $n_0(T)$ and of the width of $n^{\star}({\bf k})$ are determined. The $n_0(T)$ can be represented by $n_0(T) = n_0(0) [1 - (T/T_\lambda )^\gamma]$ with a small $n_0(0) = 2.80 \pm 0.20$ \% and large $\gamma = 13\pm 2$ for $T < T_\lambda$ indicating strong interaction. The onset of BEC is accompanied by a significant narrowing of the $n^\star({\bf k})$. The narrowing accounts for 65 \% of the drop in kinetic energy below $T_{\lambda}$ and reveals an important coupling between BEC and $k >0$ states.  The experimental results are well reproduced by Path Integral Monte Carlo calculations.  
\end{abstract}
\maketitle

Bose-Einstein condensation (BEC) is pervasive in condensed matter and the origin of spectacular properties \cite{Griffin:95}. BEC may be defined as the condensation of a macroscopic fraction of Bosons into one single  particle state\cite{Einstein:24,Leggett:01}, as the onset of long range order in the one-body density  matrix\cite{Penrose:56} or in a pair function. The phase of the macroscopically occupied single particle state or pair  function introduces phase coherence in the system which is the origin of superfluidity and superconductivity. Magnetic order is also regularly described\cite{Giamarchi:08} in terms of condensation. BEC in a gas of photons has been observed\cite{Klaers:10}. Particularly, remarkable properties in dilute gases in traps arise from BEC and superflow. In gases the fraction, $n_0$, of Bosons in the condensate can be 100 \% and BEC is easier to observe than superflow. In contrast, in dense systems such as liquid $^4$He, where $n_0$ is small, superflow was observed long before BEC \cite{Nozieres:90}. To date, BEC is uniquely observed in helium in the dynamic structure factor using neutrons\cite{Silver:89a,Glyde:94,Mayers:97b,Andreani:05}.

Reports of possible superflow in solid helium\cite{Kim:04a,Kim:04,Choi:10,Pratt:11,Balibar:10} have stimulated renewed 
interest in BEC in dense Bose systems. Observation of BEC in solid helium would be an unambiguous verification of superflow but, as yet, has not been observed\cite{Diallo:07,Adams:07,Diallo:09}. To better understand BEC in dense systems we have measured\cite{Glyde:11a} the condensate fraction in liquid \4he~ at low temperature as a function of pressure up to solidification, $p$ = 25.3 bar. The full atomic momentum distribution, $n({\bf k})$, and especially the impact of BEC on $n({\bf k})$ in dense systems, is also of great interest.

In this Letter we report precision measurements of the temperature dependence of $n({\bf k})$ and $n_0$ of liquid \4he under pressure $p$ = 24 bar. The measurements were made on the ARCS instrument at the Spallation Neutron Source (SNS), Oak Ridge National Laboratory (ORNL). Path integral Monte Carlo (PIMC) calculations are also reported. From the observed $n({\bf k})$ we obtain a Bose-Einstein condensate fraction $n_0(T) = n_0(0) [1 - (T/T_\lambda )^\gamma]$  with $n_0(0) = 2.80 \pm 0.20$\% and $\gamma = 13 \pm 2$ below the normal-superfluid transition temperature \tl = 1.86 K. The small value of \no~ and the large value of $\gamma$ signal strong interaction in the liquid at 24 bar. In addition to BEC, the momentum distribution of the atoms above the condensate, denoted \nstark, narrows below \tl. With the improved precision on ARCS, we are able to determine both the temperature dependence of \noft~ and the width of \nstark~ simultaneously. The temperature dependence of the width of \nstark~below \tl~ tracks \noft. This signals a coupling between BEC and the occupation of the higher momentum states.  Below \tl, there is both BEC and a re-distribution of occupation of the $k~> 0$ states in an interacting Bose system.

\begin{figure}
\includegraphics[scale=0.45]{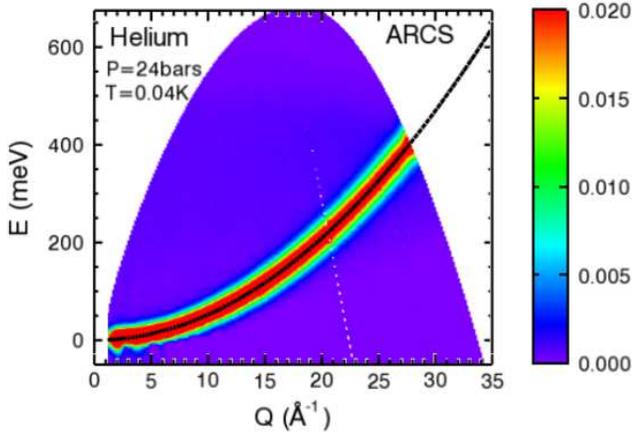}
\caption{Observed scattering intensity $S(Q,\omega)$ as a function of energy transfer $E=\hbar\omega$ and momentum transfer $\hbar Q$  from liquid $^4$He at $p=24$ bar and $T=40$ mK. Signal from the empty Al container has been subtracted. The dashed line is the calculated $^4$He recoil line, $E_r=\hbar^2 Q^2/2m$, shown as a guide to the eye.}
\label{Fig1}
\end{figure}

At temperatures close to, but above $T_{\lambda}$, the kinetic energy, $\langle K\rangle$, is dominated by quantum zero-point motion and hence is relatively temperature independent. When the liquid is cooled below \tl, the kinetic energy, \ke, drops precipitously both due to the onset of BEC and as a result of a narrowing of \nstark~ with temperature. The present results show that at  24 bar  approximately 65\%  of the observed drop in $\langle K\rangle$ comes from the  decrease in the width of \nstark~ while 35\% arises from the onset of BEC. This means that determinations of the condensate fraction, \no, from the drop in \ke~ below \tl~ must take account of this narrowing of \nstark. Otherwise \no~ will be overestimated. The impact of the narrowing is relatively smaller at saturated vapor pressure (SVP) where \no~ is larger. However, this effect explains why determinations of \no~ from the \ke~ made assuming a change in weight, but no change in the shape, of \nstark~ with temperature yield large values of  \no\cite{Mayers:97b,Mayers:00}. The temperature dependence of both \noft~ and the width of \nstark~ are well reproduced in PIMC calculations.

The atomic momentum distribution is observed in the dynamic structure factor, $S(Q,\omega)$, at high momentum, $\hbar Q$, and energy, $\hbar \omega$, transfer. In this limit, denoted the impulse approximation (IA), the energy transfer to the sample by the scattered neutrons is quadratic in $Q$, and centered around the $^4$He recoil line $E_r=\hbar^2 Q^2/2m$, as shown in Fig.\ref{Fig1}.  In the IA, $S(Q,\omega)$ is conveniently expressed in terms of the $y$-scaling variable, $y$=$(\omega-\omega_r)/v_r$, yielding\cite{Silver:89a,Glyde:94,Andreani:05},
\begin{equation}
J_{IA}(y)=v_r S(Q,\omega)=\int d{\bf k} \delta(y-{k}_Q) n({\bf k}),
\label{eq_jia}
\end{equation}
where $k_Q=k.\frac{\bf Q}{|Q|}$ and $v_r=\hbar Q/m$. $J_{IA}(y)$ is denoted the longitudinal momentum distribution and its Fourier transform, $J_{IA}(s)$, given by $J_{IA}(s)=\int_{-\infty}^{+\infty} J_{IA}(y)e^{-iys}ds$ is the one body density matrix (OBDM) for displacements $s={\bf r}.\hat{\bf Q}$ along ${\bf Q}$. At finite $Q$, the observed $J(Q,y)$ is broadened by final state interactions \cite{Glyde:94a} and the instrument resolution.  Accounting for these effects, single particles dynamics such as $n_0$, and $n({\bf k})$ are directly observed from $J(Q,y)$.

\begin{figure}
\includegraphics[scale=0.4]{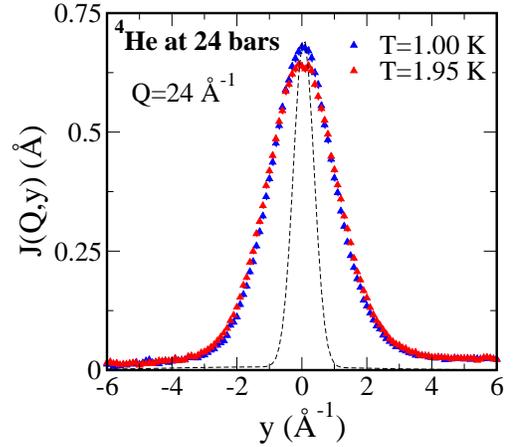}
\caption{Observed $J(Q,y)$ at $Q=24$ {\AA$^{-1}$} and at the temperatures indicated. The dashed line is the measured ARCS resolution function at $Q=24$ {\AA$^{-1}$}. The increased peak height at low temperature is attributed to the onset of BEC.}
\label{Fig3}
\end{figure}

The ARCS instrument was set in its high resolution  mode and a neutron incident energy $E_i=700$ meV was selected to allow access to wavevectors up to $Q=28$ {\AA}${^{-1}}$. Fig.~\ref{Fig1} displays the net 2D contour map obtained from liquid $^4$He after background subtraction. Fig. \ref{Fig3}  shows the observed $J(Q,y)$ of liquid $^4$He at $Q=24$ {\AA}${^{-1}}$ at temperatures below and above $T_\lambda$ along with the instrument resolution function. The relative increased in intensity at $y=0$ as the temperature is lowered below $T_\lambda$ is attributed to the onset of BEC.

To analyze the data, we follow methods tested previously in liquid $^4$He at SVP \cite{Sears:84,Reiter:85,Silver:89a,Glyde:94a,Azuah:00,Andreani:05}. Specifically, we express  $J(Q,s)$ as a product of the ideal $J_{IA}(s)$, and the final state (FS) function $R(Q,s)$ \cite{Glyde:94a,Azuah:00}. $J_{IA}(s)$ and $R(Q,s)$ are determined separately from fits to data.  In order to extract a condensate $n_0$,  we assumed as in previous work \cite{Glyde:94a,Azuah:00} a model momentum distribution $n({\bf k})$ of the form, 
\begin{equation}
n({\bf k})=n_0[\delta({\bf k})+f({\bf k})]+A_1n^\star({\bf k}),
\label{eq_nk}
\end{equation}
where $n_0\delta({\bf k})$ is the condensate component, $n^\star({\bf k})$ is the distribution of atoms above the condensate in the $k\neq0$ states and $n_0f({\bf k})$ a coupling between the two.  The Fourier transform of $n^\star({\bf k})$, $J^\star_{IA}(s)=n^\star(s)$,  expanded in powers of $s$ up to $s^6$ is,
\begin{equation}
n^\star(s)=\exp\left[-\frac{\bar\alpha_2s^2}{2!}+\frac{\bar\alpha_4s^4}{4!}-\frac{\bar\alpha_6s^6}{6!}\right]\
\label{eq_ns}
\end{equation}

\noindent The model $n({\bf k})$ in Eq.  \ref{eq_nk} has thus four adjustable parameters, $n_0$, $\bar\alpha_2$, $\bar\alpha_4$,  and $\bar\alpha_6$ that can be obtained by fits to experimental data. Including resolution effects, we were able to determine $n_0(T)$, the momentum distribution $n^\star({\bf k})$ and the final state function $R(Q,y)$.  

\begin{figure}
\includegraphics[scale=0.45]{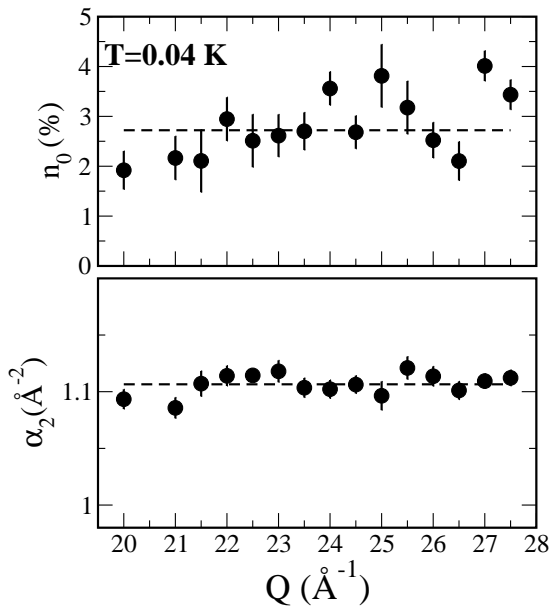}
\caption{The parameters $\bar\alpha_2$ and $n_0$ obtained by fits to data at several $Q$ values and temperature $T=40$ mK. }
\label{Fig4}
\end{figure}

To get a microscopic understanding of our data, we have carried out Path Integral Monte Carlo (PIMC) calculations of the momentum distribution liquid $^4$He at the same densities and temperatures  covered by the experiment \cite{Wilks:67}.  PIMC is a microscopic stochastic method that is able to generate very accurate results relying only on the Hamiltonian of the system. The results here presented are obtained using a well tested Aziz potential and a modern approach based on a high-order action and the worm algorithm for a better sampling of permutations \cite{Rota:11a}.  

\begin{table}
 \caption{Temperature dependence of the condensate fraction $n_0$ and n({\bf k}) parameters in liquid $^4$He under pressure $p=$24 bars . The $\lambda$ transition is at $T=$1.86 K. The same parameters in liquid $^4$He at SVP \cite{Azuah:00} are shown for comparison.}
  \begin{tabular}{ |c |c |c |c |c |c |}
  \hline
  \hline
   P (bar)&T (K) & $n_0$ (\%) &  $\bar\alpha_2$ (\AA${^{-2}}$) &  $\bar\alpha_4$ (\AA$^{{-4}}$) & $\bar\alpha_6$(\AA$^{{-6}}$) \\ 
    \hline
 24    & 0.04  & 2.88$\pm$0.60 & 1.10$\pm$0.02 & 0.63$\pm$0.10 &1.35$\pm$0.20 \\ 
          & 1.00  & 2.96$\pm$0.70 & 1.11$\pm$0.02 & 0.62$\pm$0.10 &1.34$\pm$0.15\\ 
          & 1.30  &  2.64$\pm$0.75 &1.11$\pm$0.02 & 0.63$\pm$0.10 & 1.28$\pm$0.15\\ 
          & 1.50  &  2.56$\pm$0.50 &1.12$\pm$0.01 & 0.61$\pm$0.10& 1.26$\pm$0.25\\ 
          & 1.75  &  1.72$\pm$0.70 & 1.15$\pm$0.02& 0.61$\pm$0.10 & 1.27$\pm$0.25\\
\hline 
 24   & 1.95 & 0.25$\pm$1.65 & 1.19$\pm$0.02   & 0.65$\pm$0.20 & 0.95$\pm$0.35 \\ 
         & 3.50 & 0.32$\pm$1.00  & 1.18$\pm$0.02  & 0.56$\pm$0.15 & 0.98$\pm$0.30\\ 
         & 5.00 & -0.04$\pm$1.20 & 1.18$\pm$0.02  & 0.53$\pm$0.20 & 0.44$\pm$0.60 \\ 
\hline
   SVP  & 0.50 & 7.25$\pm$0.75 & 0.897$\pm$0.02 & 0.46$\pm$0.05 & 0.38$\pm$0.04\\
   \hline
   \hline
    \end{tabular}
\label{tbl1}
    \end{table}
    
\begin{figure}
\includegraphics[scale=0.325]{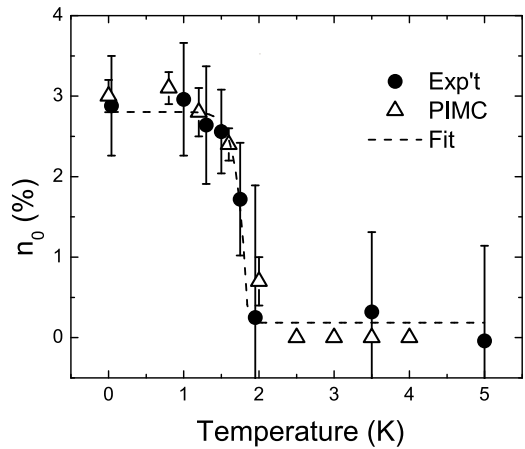}
\caption{Observed condensate fraction as a function of temperature. The dashed line is a line fit to the experimental data using of $n_0(T)=n_0(0)[1-(T/T_\lambda)^\gamma]$, where $T_{\lambda}=1.86$ K. The open triangles are  the simulated PIMC results.}
\label{Fig5}
\end{figure}

\begin{figure}
\includegraphics[scale=0.325]{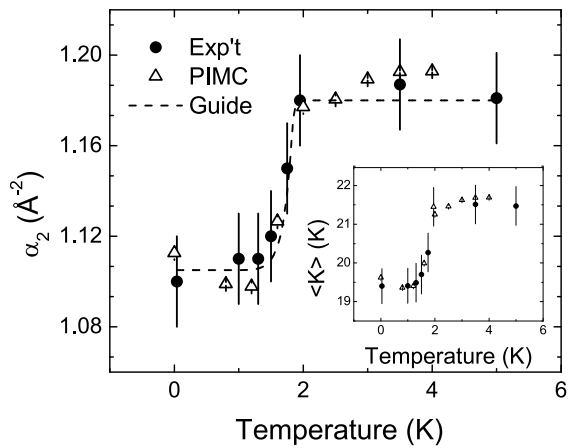}
\caption{Temperature dependence of the  width $\bar\alpha_2$ of $n^\star({\bf k})$ at $p=$24 bar: simulation and experiment. The dashed line shows that $\bar\alpha_2(T)$ has a temperature dependence that tracks $n_0(T)$. The inset shows the corresponding $\langle K\rangle(T)$.}
\label{Fig6}
\end{figure}

Fig. \ref{Fig4} shows the  parameters $n_0$ and $\bar\alpha_2$  obtained by fits to the experimental data at several $Q$ values and at $T=40$ mK. The variation with $Q$ arises from the statistical uncertainty of the data.  The dashed lines indicate the corresponding average values. The average condensate fraction, \no, and the parameters \alt, \alf, and \als~ obtained by fits to data are listed in Table \ref{tbl1}.  The dependence of \noft~ and \alt(T) on temperature is shown in Figs. \ref{Fig5} and \ref{Fig6}, respectively. From Table \ref{tbl1} and Fig. \ref{Fig5}, we see that \no~ reaches a maximum value of \no~= 2.88 $\pm$ 0.60 \% at low temperature. As the temperature is decreased below \tl~ = 1.86 K, \noft~ increases rapidly toward its maximum value. Essentially, \noft~ plateaus to its maximum value at temperatures close to \tl~ = 1.86 K. This indicates that strong interaction between the \4he atoms limits \no~ at higher pressure with a decrease to the lowest temperatures unable to reduce \no~ further.  While the values of the parameters $\bar\alpha_4$ and $\bar\alpha_6$ in Table \ref{tbl1} are somewhat higher than those observed previously  \cite{Glyde:11a} at low temperature, the total low temperature $n^\star({\bf k})$ is the same and the value of $n_0$ at low $T$ is independent of which $n^\star({\bf k})$ is used.

In a Bose gas, \noft~ = \no(0)[1 - (T/\tl)$^\gamma$] with \no~ = 100 \% and $\gamma$ = 3/2. A fit of this expression to the observed \noft~ in liquid \4he at SVP \cite{Azuah:00} gives \no(0) = 7.25 $\pm$ 0.75 \% and   $\gamma$ = 5.5 $\pm$ 1.0. A fit of the same expression to the present observed \noft~ at 24 bar  gives \no(0) = 2.8 $\pm$ 0.20 \% and $\gamma$ = 13 $\pm$ 2.0. The fit is shown as a dashed line in Fig. \ref{Fig5}.  The large value of gamma reflects the strong interaction in liquid $^4$He at 24 bar.

From Table \ref{tbl1} and Fig. \ref{Fig6}, we see that the parameter \alt~= $\langle |k_Q|^2\rangle$ which sets the width of \nstark~ decreases from 1.18 {\AA}${^{-2}}$ in the normal phase ($T>$~\tl) to 1.10 -1.11 {\AA}${^{-2}}$ at low temperature. That is, while \alt~ is approximately independent of $T$ in the normal phase, \alt~ drops abruptly at temperatures immediately below \tl. This abrupt decrease is unlikely to be a thermal effect since the thermal energy $k_B$\tl~ is already much less than the zero point energy (approximately \ke~ = 21.47 K). Rather, the abrupt drop of \alt~ below \tl~ suggests a link to the onset of BEC. To test this picture we show the function \alt(T) = \alt(\tl) - $\Delta$ [1 - (T/\tl)$^\gamma$]  where \alt(\tl) = 1.18 {\AA}${^{-2}}$, $\Delta$ = \alt(\tl) - \alt(0) = 0.075 {\AA}$^{{-2}}$ and $\gamma=13$ as a dashed line in Fig. \ref{Fig6} which has the same temperature dependence as $n_0(T)$.  The dashed line reproduces the observed $\bar\alpha_2(T)$ well.  Below \tl, there appears to be a coupling between \no~ and \nstark, perhaps of the same form as \no $f({\bf k})$, which leads to a narrowing of \nstark.

The narrowing of \nstark~ below \tl~ is reproduced by PIMC calculations. The present PIMC values of \alt~ are shown in Fig. \ref{Fig6} and they also decrease abruptly below \tl. Thus the sharp reduction of \alt~ below \tl, not observed previously but observable with the increased precision of the ARCS neutron scattering instrument, is supported by accurate PIMC calculations. Below \tl~ there is both BEC and a narrowing of \nstark~ in a strongly interacting Bose liquid.           

In summary, the observed and PIMC values of $n_0(T)$~ in liquid $^4$He at 24 bar near the solidification line ($p$ = 25.3 bar) are well represented by $n_0(T) = n_0(0) [1 - (T/T_\lambda )^\gamma]$ with $n_0(0) = 2.80 \pm 0.20$ \% and $\gamma = 13 \pm 2$. In a Bose gas $\gamma$ = 1.5 and in liquid $^4$He at SVP $\gamma = 5.5 \pm 1.0$. The large $\gamma$~ at 24 bar indicates strong interaction in the liquid with $n_0$ saturating to a small value at temperatures close to $T_\lambda$. On cooling below $T_\lambda$, there is both BEC and a narrowing of the atomic momentum distribution of the atoms above the condensate, $n^\star({\bf k})$. The narrowing is characterized here by a drop in the width, $\bar\alpha_2$ = $\langle |k_Q|^2 \rangle$, of $n^\star({\bf k})$. The temperature dependence of $\bar\alpha_2$ below $T_\lambda$ tracks $n_0(T)$ indicating that interaction  between  the condensate and the higher momentum states causes the narrowing of $n^\star({\bf k})$. This coupling between $n_0$ and $n^\star({\bf k})$ is currently not understood.

Both BEC and the narrowing of   $n^\star({\bf k})$ contribute to the drop in the $\langle K\rangle$ at temperatures below $T_\lambda$. The present observed and PIMC values of the temperature dependence of $n_0(T)$ and $\langle K\rangle$ at 24 bar agree well.  At $p$ = 24 bar, approximately 65\% of the drop in $\langle K\rangle$ arises from the narrowing of $n^{\star}({\bf k})$ below $T_\lambda$. Thus an $n_0$ obtained from the $\langle K\rangle$ assuming no narrowing of $n^\star({\bf k})$ would significantly overestimate $n_0$.  Indeed, if we apply the method to the present data, we get $n_0$=7.5\% at 40 mK, which is more than 2 times the observed value. In the liquid at  SVP where $n_0$ is larger, the relative reduction of the $\langle K\rangle$ arising from the narrowing of $n^\star({\bf k})$ is smaller. However, it is still significant and existing values of $n_0$ at SVP determined from the $\langle K\rangle$ assuming no narrowing of \nstark~ may have to be corrected.

We thank J. Carmichael for designing the modular 100 bar sample cell  and L. Solomon and E. Robles for valuable technical assistance with the sample environment. R. R. and J. B. acknowledge partial financial support from the DGI (Spain) Grant No.~FIS2008-04403 and Generalitat de Catalunya Grant No.~2009SGR-1003. Work at ORNL and SNS is sponsored by the Scientific User Facilities Division, Office of Basic Energy Sciences, US DOE. This work was supported by the DOE, Office of Basic Energy Sciences under contract No. ER46680.


\begin{thebibliography}{27}
\expandafter\ifx\csname natexlab\endcsname\relax\def\natexlab#1{#1}\fi
\expandafter\ifx\csname bibnamefont\endcsname\relax
  \def\bibnamefont#1{#1}\fi
\expandafter\ifx\csname bibfnamefont\endcsname\relax
  \def\bibfnamefont#1{#1}\fi
\expandafter\ifx\csname citenamefont\endcsname\relax
  \def\citenamefont#1{#1}\fi
\expandafter\ifx\csname url\endcsname\relax
  \def\url#1{\texttt{#1}}\fi
\expandafter\ifx\csname urlprefix\endcsname\relax\def\urlprefix{URL }\fi
\providecommand{\bibinfo}[2]{#2}
\providecommand{\eprint}[2][]{\url{#2}}

\bibitem[{\citenamefont{Griffin et~al.}(1995)\citenamefont{Griffin, Snoke, and
  Stringari}}]{Griffin:95}
\bibinfo{editor}{\bibfnamefont{A.}~\bibnamefont{Griffin}},
  \bibinfo{editor}{\bibfnamefont{D.}~\bibnamefont{Snoke}}, \bibnamefont{and}
  \bibinfo{editor}{\bibfnamefont{S.}~\bibnamefont{Stringari}}, eds.,
  \emph{\bibinfo{title}{{Bose-Einstein Condensation}}}
  (\bibinfo{publisher}{Cambridge University Press},
  \bibinfo{address}{Cambridge, England}, \bibinfo{year}{1995}).

\bibitem[{\citenamefont{Einstein}(1924)}]{Einstein:24}
\bibinfo{author}{\bibfnamefont{A.}~\bibnamefont{Einstein}},
  \bibinfo{journal}{Sitzungsber. Kgl. Preuss. Akad. Wiss., Phys. Math. Kl.}
  \textbf{\bibinfo{volume}{261}} (\bibinfo{year}{1924}).

\bibitem[{\citenamefont{Leggett}(2001)}]{Leggett:01}
\bibinfo{author}{\bibfnamefont{A.~J.} \bibnamefont{Leggett}},
  \bibinfo{journal}{Rev. Mod. Phys.} \textbf{\bibinfo{volume}{73}},
  \bibinfo{pages}{307} (\bibinfo{year}{2001}).

\bibitem[{\citenamefont{Penrose and Onsager}(1956)}]{Penrose:56}
\bibinfo{author}{\bibfnamefont{O.}~\bibnamefont{Penrose}} \bibnamefont{and}
  \bibinfo{author}{\bibfnamefont{L.}~\bibnamefont{Onsager}},
  \bibinfo{journal}{Phys. Rev.} \textbf{\bibinfo{volume}{104}},
  \bibinfo{pages}{576} (\bibinfo{year}{1956}).

\bibitem[{\citenamefont{Giamarchi et~al.}(2008)\citenamefont{Giamarchi, Ruegg,
  and Tchernyshyov}}]{Giamarchi:08}
\bibinfo{author}{\bibfnamefont{T.}~\bibnamefont{Giamarchi}},
  \bibinfo{author}{\bibfnamefont{G.}~\bibnamefont{Ruegg}}, \bibnamefont{and}
  \bibinfo{author}{\bibfnamefont{O.}~\bibnamefont{Tchernyshyov}},
  \bibinfo{journal}{{Nat. Phys.}} \textbf{\bibinfo{volume}{4}},
  \bibinfo{pages}{198} (\bibinfo{year}{2008}).

\bibitem[{\citenamefont{Klaers et~al.}(2010)\citenamefont{Klaers, Schmitt,
  Vewinger, and Weitz}}]{Klaers:10}
\bibinfo{author}{\bibfnamefont{J.}~\bibnamefont{Klaers}},
  \bibinfo{author}{\bibfnamefont{J.}~\bibnamefont{Schmitt}},
  \bibinfo{author}{\bibfnamefont{F.}~\bibnamefont{Vewinger}}, \bibnamefont{and}
  \bibinfo{author}{\bibfnamefont{M.}~\bibnamefont{Weitz}},
  \bibinfo{journal}{Nature} \textbf{\bibinfo{volume}{468}},
  \bibinfo{pages}{545} (\bibinfo{year}{2010}).

\bibitem[{\citenamefont{Nozi\`{e}res and Pines}(1990)}]{Nozieres:90}
\bibinfo{author}{\bibfnamefont{P.}~\bibnamefont{Nozi\`{e}res}}
  \bibnamefont{and} \bibinfo{author}{\bibfnamefont{D.}~\bibnamefont{Pines}},
  \emph{\bibinfo{title}{{Theory of Quantum Liquids Vol. II}}}
  (\bibinfo{publisher}{Addison-Wesley}, \bibinfo{address}{Redwood City, CA},
  \bibinfo{year}{1990}).

\bibitem[{\citenamefont{Glyde}(1994{\natexlab{a}})}]{Glyde:94}
\bibinfo{author}{\bibfnamefont{H.~R.} \bibnamefont{Glyde}},
  \emph{\bibinfo{title}{{Excitations in Liquid and Solid Helium}}}
  (\bibinfo{publisher}{Oxford University Press}, \bibinfo{address}{Oxford,
  England}, \bibinfo{year}{1994}{\natexlab{a}}).

\bibitem[{\citenamefont{Andreani et~al.}(2005)\citenamefont{Andreani,
  Colognesi, Mayers, Reiter, and Senesi}}]{Andreani:05}
\bibinfo{author}{\bibfnamefont{C.}~\bibnamefont{Andreani}},
  \bibinfo{author}{\bibfnamefont{D.}~\bibnamefont{Colognesi}},
  \bibinfo{author}{\bibfnamefont{J.}~\bibnamefont{Mayers}},
  \bibinfo{author}{\bibfnamefont{G.~F.} \bibnamefont{Reiter}},
  \bibnamefont{and} \bibinfo{author}{\bibfnamefont{R.}~\bibnamefont{Senesi}},
  \bibinfo{journal}{Adv. Phys.} \textbf{\bibinfo{volume}{54}},
  \bibinfo{pages}{377} (\bibinfo{year}{2005}).

\bibitem[{\citenamefont{Mayers et~al.}(1997)\citenamefont{Mayers, Andreani, and
  Colognesi}}]{Mayers:97b}
\bibinfo{author}{\bibfnamefont{J.}~\bibnamefont{Mayers}},
  \bibinfo{author}{\bibfnamefont{C.}~\bibnamefont{Andreani}}, \bibnamefont{and}
  \bibinfo{author}{\bibfnamefont{D.}~\bibnamefont{Colognesi}},
  \bibinfo{journal}{J. Phys. Condens. Mat.} \textbf{\bibinfo{volume}{9}},
  \bibinfo{pages}{10639} (\bibinfo{year}{1997}).

\bibitem[{\citenamefont{Silver and Sokol}(1989)}]{Silver:89a}
\bibinfo{author}{\bibfnamefont{R.~N.} \bibnamefont{Silver}} \bibnamefont{and}
  \bibinfo{author}{\bibfnamefont{P.~E.} \bibnamefont{Sokol}},
  \emph{\bibinfo{title}{{Momentum Distributions}}}
  (\bibinfo{publisher}{Plenum}, \bibinfo{address}{New York},
  \bibinfo{year}{1989}).

\bibitem[{\citenamefont{Kim and Chan}(2004{\natexlab{a}})}]{Kim:04a}
\bibinfo{author}{\bibfnamefont{E.}~\bibnamefont{Kim}} \bibnamefont{and}
  \bibinfo{author}{\bibfnamefont{M.~H.~W.} \bibnamefont{Chan}},
  \bibinfo{journal}{Nature (London)} \textbf{\bibinfo{volume}{427}},
  \bibinfo{pages}{225} (\bibinfo{year}{2004}{\natexlab{a}}).

\bibitem[{\citenamefont{Kim and Chan}(2004{\natexlab{b}})}]{Kim:04}
\bibinfo{author}{\bibfnamefont{E.}~\bibnamefont{Kim}} \bibnamefont{and}
  \bibinfo{author}{\bibfnamefont{M.~H.~W.} \bibnamefont{Chan}},
  \bibinfo{journal}{Science} \textbf{\bibinfo{volume}{305}},
  \bibinfo{pages}{1941} (\bibinfo{year}{2004}{\natexlab{b}}).

\bibitem[{\citenamefont{Balibar}(2010)}]{Balibar:10}
\bibinfo{author}{\bibfnamefont{S.}~\bibnamefont{Balibar}},
  \bibinfo{journal}{Nature (London)} \textbf{\bibinfo{volume}{464}},
  \bibinfo{pages}{176} (\bibinfo{year}{2010}).

\bibitem[{\citenamefont{Choi et~al.}(2010)\citenamefont{Choi, Takahashi, Kono,
  and Kim}}]{Choi:10}
\bibinfo{author}{\bibfnamefont{H.}~\bibnamefont{Choi}},
  \bibinfo{author}{\bibfnamefont{D.}~\bibnamefont{Takahashi}},
  \bibinfo{author}{\bibfnamefont{K.}~\bibnamefont{Kono}}, \bibnamefont{and}
  \bibinfo{author}{\bibfnamefont{E.}~\bibnamefont{Kim}},
  \bibinfo{journal}{Science} \textbf{\bibinfo{volume}{330}},
  \bibinfo{pages}{1512} (\bibinfo{year}{2010}).

\bibitem[{\citenamefont{Pratt et~al.}(2011)\citenamefont{Pratt, Hunt, Gadagkar,
  Yamashita, Graf, Balatsky, and Davis}}]{Pratt:11}
\bibinfo{author}{\bibfnamefont{E.~J.} \bibnamefont{Pratt}},
  \bibinfo{author}{\bibfnamefont{B.}~\bibnamefont{Hunt}},
  \bibinfo{author}{\bibfnamefont{V.}~\bibnamefont{Gadagkar}},
  \bibinfo{author}{\bibfnamefont{M.}~\bibnamefont{Yamashita}},
  \bibinfo{author}{\bibfnamefont{M.~J.} \bibnamefont{Graf}},
  \bibinfo{author}{\bibfnamefont{A.~V.} \bibnamefont{Balatsky}},
  \bibnamefont{and} \bibinfo{author}{\bibfnamefont{J.~C.} \bibnamefont{Davis}},
  \bibinfo{journal}{Science} \textbf{\bibinfo{volume}{332}},
  \bibinfo{pages}{821} (\bibinfo{year}{2011}).

\bibitem[{\citenamefont{Diallo et~al.}(2007)\citenamefont{Diallo, Pearce,
  Azuah, Kirichek, Taylor, and Glyde}}]{Diallo:07}
\bibinfo{author}{\bibfnamefont{S.~O.} \bibnamefont{Diallo {\em et al.}}},
  \bibinfo{journal}{Phys. Rev. Lett.} \textbf{\bibinfo{volume}{98}},
  \bibinfo{pages}{205301} (\bibinfo{year}{2007}).

\bibitem[{\citenamefont{Diallo et~al.}(2009)\citenamefont{Diallo, Azuah,
  Kirichek, Taylor, and Glyde}}]{Diallo:09}
\bibinfo{author}{\bibfnamefont{S.~O.} \bibnamefont{Diallo {\em et al.}}},
  \bibinfo{journal}{Phys. Rev. B} \textbf{\bibinfo{volume}{80}},
  \bibinfo{pages}{060504} (\bibinfo{year}{2009}).

\bibitem[{\citenamefont{Adams et~al.}(2007)\citenamefont{Adams, Mayers,
  Kirichek, and Down}}]{Adams:07}
\bibinfo{author}{\bibfnamefont{M.~A.} \bibnamefont{Adams}},
  \bibinfo{author}{\bibfnamefont{J.}~\bibnamefont{Mayers}},
  \bibinfo{author}{\bibfnamefont{O.}~\bibnamefont{Kirichek}}, \bibnamefont{and}
  \bibinfo{author}{\bibfnamefont{R.~B.~E.} \bibnamefont{Down}},
  \bibinfo{journal}{Phys. Rev. Lett.} \textbf{\bibinfo{volume}{98}},
  \bibinfo{pages}{085301} (\bibinfo{year}{2007}).

\bibitem[{\citenamefont{Glyde et~al.}(2011)\citenamefont{Glyde, Diallo, Azuah,
  Kirichek, and Taylor}}]{Glyde:11a}
\bibinfo{author}{\bibfnamefont{H.~R.} \bibnamefont{Glyde}},
  \bibinfo{author}{\bibfnamefont{S.~O.} \bibnamefont{Diallo}},
  \bibinfo{author}{\bibfnamefont{R.~T.} \bibnamefont{Azuah}},
  \bibinfo{author}{\bibfnamefont{O.}~\bibnamefont{Kirichek}}, \bibnamefont{and}
  \bibinfo{author}{\bibfnamefont{J.~W.} \bibnamefont{Taylor}},
  \bibinfo{journal}{Phys. Rev. B} \textbf{\bibinfo{volume}{83}},
  \bibinfo{pages}{100507 (R)} (\bibinfo{year}{2011}).

\bibitem[{\citenamefont{Mayers et~al.}(2000)\citenamefont{Mayers, Albergamo,
  and Timms}}]{Mayers:00}
\bibinfo{author}{\bibfnamefont{J.}~\bibnamefont{Mayers}},
  \bibinfo{author}{\bibfnamefont{F.}~\bibnamefont{Albergamo}},
  \bibnamefont{and} \bibinfo{author}{\bibfnamefont{D.}~\bibnamefont{Timms}},
  \bibinfo{journal}{Physica B: Condensed Matter}
  \textbf{\bibinfo{volume}{276-278}}, \bibinfo{pages}{811}
  (\bibinfo{year}{2000}).

\bibitem[{\citenamefont{Glyde}(1994{\natexlab{b}})}]{Glyde:94a}
\bibinfo{author}{\bibfnamefont{H.~R.} \bibnamefont{Glyde}},
  \bibinfo{journal}{Phys. Rev. B} \textbf{\bibinfo{volume}{50}},
  \bibinfo{pages}{6726} (\bibinfo{year}{1994}{\natexlab{b}}).

\bibitem[{\citenamefont{Sears}(1984)}]{Sears:84}
\bibinfo{author}{\bibfnamefont{V.~F.} \bibnamefont{Sears}},
  \bibinfo{journal}{Phys. Rev. B} \textbf{\bibinfo{volume}{30}},
  \bibinfo{pages}{44} (\bibinfo{year}{1984}).

\bibitem[{\citenamefont{Reiter and Silver}(1985)}]{Reiter:85}
\bibinfo{author}{\bibfnamefont{G.}~\bibnamefont{Reiter}} \bibnamefont{and}
  \bibinfo{author}{\bibfnamefont{R.}~\bibnamefont{Silver}},
  \bibinfo{journal}{Phys. Rev. Lett.} \textbf{\bibinfo{volume}{54}},
  \bibinfo{pages}{1047} (\bibinfo{year}{1985}).

\bibitem[{\citenamefont{Glyde et~al.}(2000)\citenamefont{Glyde, Azuah, and
  Stirling}}]{Azuah:00}
\bibinfo{author}{\bibfnamefont{H.~R.} \bibnamefont{Glyde}},
  \bibinfo{author}{\bibfnamefont{R.~T.} \bibnamefont{Azuah}}, \bibnamefont{and}
  \bibinfo{author}{\bibfnamefont{W.~G.} \bibnamefont{Stirling}},
  \bibinfo{journal}{Phys. Rev. B} \textbf{\bibinfo{volume}{62}},
  \bibinfo{pages}{14337} (\bibinfo{year}{2000}).

\bibitem[{\citenamefont{Wilks}(1967)}]{Wilks:67}
\bibinfo{author}{\bibfnamefont{J.}~\bibnamefont{Wilks}},
  \emph{\bibinfo{title}{The Properties of Liquid and Solid Helium}}
  (\bibinfo{publisher}{Oxford: Clarendon Press}, \bibinfo{year}{1967}).

\bibitem[{\citenamefont{Rota and Boronat}(2011)}]{Rota:11a}
\bibinfo{author}{\bibfnamefont{R.}~\bibnamefont{Rota}} \bibnamefont{and}
  \bibinfo{author}{\bibfnamefont{J.}~\bibnamefont{Boronat}},
  \bibinfo{journal}{J. Low Temp. Phys.} \textbf{\bibinfo{volume}{162}},
  \bibinfo{pages}{146} (\bibinfo{year}{2011}).

\end{thebibliography}



\end{document}